\documentclass[12pt,preprint]{aastex62}
\usepackage{natbib}
\usepackage{graphicx,amsmath,amssymb,gensymb}
\usepackage{url}

\newcommand{\bdv}[1]{\mbox{\boldmath$#1$}}

\def\au{{\rm AU}} 
 
\def\kms{{\rm km}\,{\rm s}^{-1}}
\def\masyr{{\rm mas}\,{\rm yr}^{-1}}

\def\mas{{\rm mas}}
\def\muas{\mu{\rm as}}

\def\max{{\rm max}}
\def\rel{{\rm rel}}

\def\geo{{\rm geo}}
\def\e{{\rm E}}
\def\bpi{{\bdv\pi}}
\def\bmu{{\bdv\mu}}

\def\btheta{{\bdv\theta}}

\begin{document}
\title{\large {\it Spitzer} Microlensing Parallax for OGLE-2017-BLG-0896 Reveals
a Counter-Rotating Low-Mass Brown Dwarf}

\author{Yossi Shvartzvald}
\affiliation{IPAC, Mail Code 100-22, Caltech, 1200 E. California Blvd., Pasadena, CA 91125, USA}

\author{Jennifer~C.~Yee}
\affiliation{Harvard-Smithsonian Center for Astrophysics, 60 Garden St.,Cambridge, MA 02138, USA}

\author{Jan~Skowron}
\affiliation{Warsaw University Observatory, Al. Ujazdowskie 4, 00-478 Warszawa, Poland}

\author{Chung-Uk Lee}
\affiliation{Korea Astronomy and Space Science Institute, Daejon 34055, Republic of Korea}
\affiliation{Korea University of Science and Technology, 217 Gajeong-ro, Yuseong-gu, Daejeon 34113, Republic of Korea}

\author{Andrzej Udalski}
\affiliation{Warsaw University Observatory, Al. Ujazdowskie 4, 00-478 Warszawa, Poland}

\author{Sebastiano Calchi Novati}
\affiliation{IPAC, Mail Code 100-22, Caltech, 1200 E. California Blvd., Pasadena, CA 91125, USA}

\author{Valerio~Bozza}
\affiliation{Dipartimento di Fisica "E.R. Caianiello", Universit{\`a} di Salerno, Via Giovanni Paolo II 132, 84084, Fisciano, Italy}
\affiliation{Istituto Nazionale di Fisica Nucleare, Sezione di Napoli, Napoli, Italy}
             
\collaboration{---}
\noaffiliation

\author{Charles A. Beichman}
\affiliation{IPAC, Mail Code 100-22, Caltech, 1200 E. California Blvd., Pasadena, CA 91125, USA}

\author{Geoffery Bryden}
\affiliation{Jet Propulsion Laboratory, California Institute of Technology, 4800 Oak Grove Drive, Pasadena, CA 91109, USA}

\author{Sean Carey}
\affiliation{IPAC, Mail Code 100-22, Caltech, 1200 E. California Blvd., Pasadena, CA 91125, USA}

\author{B.~Scott~Gaudi}
\affiliation{Department of Astronomy, Ohio State University, 140 W. 18th Ave., Columbus, OH  43210, USA}

\author{Calen~B.~Henderson}
\affiliation{IPAC, Mail Code 100-22, Caltech, 1200 E. California Blvd., Pasadena, CA 91125, USA}

\author{Wei Zhu}
\affiliation{Canadian Institute for Theoretical Astrophysics, University of Toronto, 60 St George Street, Toronto, ON M5S 3H8, Canada}

\collaboration{($Spitzer$ team)}
\noaffiliation

\author{Etienne Bachelet}
\affiliation{Las Cumbres Observatory, 6740 Cortona Drive, suite 102, Goleta, CA 93117, USA}

\author{Greg Bolt}
\affiliation{Craigie, Western Australia, Australia}

\author{Grant Christie}
\affiliation{Auckland Observatory, Auckland, New Zealand}

\author{Dan Maoz}
\affiliation{School of Physics and Astronomy, Tel-Aviv University, Tel-Aviv 6997801, Israel}

\author{Tim Natusch}
\affiliation{Auckland Observatory, Auckland, New Zealand}
\affiliation{Institute for Radio Astronomy and Space Research (IRASR), AUT University, Auckland, New Zealand}

\author{Richard W. Pogge}
\affiliation{Department of Astronomy, Ohio State University, 140 W. 18th Ave., Columbus, OH  43210, USA}

\author{Rachel  A. Street}
\affiliation{Las Cumbres Observatory, 6740 Cortona Drive, suite 102, Goleta, CA 93117, USA}

\author{Thiam-Guan Tan}
\affiliation{Perth Exoplanet Survey Telescope, Perth, Australia}

\author{Yiannis Tsapras}
\affiliation{Astronomisches Rechen-Institut, Zentrum f{\"u}r Astronomie der Universit{\"a}t Heidelberg (ZAH), 69120 Heidelberg, Germany}

\collaboration{(LCO and $\mu$FUN Follow-up Teams)} 
\noaffiliation

\author{Pawe{\l} Pietrukowicz}
\affiliation{Warsaw University Observatory, Al. Ujazdowskie 4, 00-478 Warszawa, Poland}

\author{Igor Soszy\'{n}ski}
\affiliation{Warsaw University Observatory, Al. Ujazdowskie 4, 00-478 Warszawa, Poland}

\author{Micha{\l}~K.~Szyma\'{n}ski}
\affiliation{Warsaw University Observatory, Al. Ujazdowskie 4, 00-478 Warszawa, Poland}

\author{Przemek Mr\'{o}z}
\affiliation{Warsaw University Observatory, Al. Ujazdowskie 4, 00-478 Warszawa, Poland}

\author{Radoslaw~Poleski}
\affiliation{Warsaw University Observatory, Al. Ujazdowskie 4, 00-478 Warszawa, Poland}
\affiliation{Department of Astronomy, Ohio State University, 140 W. 18th Ave., Columbus, OH  43210, USA}

\author{Szymon Koz{\l}owski}
\affiliation{Warsaw University Observatory, Al. Ujazdowskie 4, 00-478 Warszawa, Poland}

\author{Krzysztof Ulaczyk}
\affiliation{Warsaw University Observatory, Al. Ujazdowskie 4, 00-478 Warszawa, Poland}

\author{Micha{\l} Pawlak}
\affiliation{Warsaw University Observatory, Al. Ujazdowskie 4, 00-478 Warszawa, Poland}

\author{Krzysztof A.~Rybicki}
\affiliation{Warsaw University Observatory, Al. Ujazdowskie 4, 00-478 Warszawa, Poland}

\author{Patryk Iwanek}
\affiliation{Warsaw University Observatory, Al. Ujazdowskie 4, 00-478 Warszawa, Poland}

\collaboration{(OGLE Collaboration)}
\noaffiliation

\author{Michael D. Albrow}
\affiliation{University of Canterbury, Department of Physics and Astronomy, Private Bag 4800, Christchurch 8020, New Zealand}

\author{Sang-Mok Cha}
\affiliation{Korea Astronomy and Space Science Institute, Daejon 34055, Republic of Korea}
\affiliation{School of Space Research, Kyung Hee University, Yongin, Kyeonggi 17104, Republic of Korea} 

\author{Sun-Ju Chung}
\affiliation{Korea Astronomy and Space Science Institute, Daejon 34055, Republic of Korea}
\affiliation{Korea University of Science and Technology, 217 Gajeong-ro, Yuseong-gu, Daejeon 34113, Republic of Korea}

\author{Andrew Gould}
\affiliation{Korea Astronomy and Space Science Institute, Daejon 34055, Republic of Korea}
\affiliation{Max-Planck-Institute for Astronomy, K\"{o}nigstuhl 17, 69117 Heidelberg, Germany}
\affiliation{Department of Astronomy, Ohio State University, 140 W. 18th Ave., Columbus, OH  43210, USA}

\author{Cheongho Han}
\affiliation{Department of Physics, Chungbuk National University, Cheongju 28644, Republic of Korea}

\author{Kyu-Ha Hwang}
\affiliation{Korea Astronomy and Space Science Institute, Daejon 34055, Republic of Korea}

\author{Youn Kil Jung}
\affiliation{Harvard-Smithsonian Center for Astrophysics, 60 Garden St.,Cambridge, MA 02138, USA}
\affiliation{Korea Astronomy and Space Science Institute, Daejon 34055, Republic of Korea}

\author{Dong-Jin Kim}
\affiliation{Korea Astronomy and Space Science Institute, Daejon 34055, Republic of Korea}

\author{Hyoun-Woo Kim}
\affiliation{Korea Astronomy and Space Science Institute, Daejon 34055, Republic of Korea}

\author{Seung-Lee Kim}
\affiliation{Korea Astronomy and Space Science Institute, Daejon 34055, Republic of Korea}
\affiliation{Korea University of Science and Technology, 217 Gajeong-ro, Yuseong-gu, Daejeon 34113, Republic of Korea}

\author{Dong-Joo Lee}
\affiliation{Korea Astronomy and Space Science Institute, Daejon 34055, Republic of Korea}

\author{Yongseok Lee}
\affiliation{Korea Astronomy and Space Science Institute, Daejon 34055, Republic of Korea}
\affiliation{School of Space Research, Kyung Hee University, Yongin, Kyeonggi 17104, Republic of Korea}

\author{Byeong-Gon Park}
\affiliation{Korea Astronomy and Space Science Institute, Daejon 34055, Republic of Korea}
\affiliation{Korea University of Science and Technology, 217 Gajeong-ro, Yuseong-gu, Daejeon 34113, Republic of Korea}

\author{Yoon-Hyun Ryu}
\affiliation{Korea Astronomy and Space Science Institute, Daejon 34055, Republic of Korea}

\author{In-Gu Shin}
\affiliation{Harvard-Smithsonian Center for Astrophysics, 60 Garden St.,Cambridge, MA 02138, USA}

\author{Weicheng Zang}
\affiliation{Physics Department and Tsinghua Centre for Astrophysics, Tsinghua University, Beijing 100084, China}
\affiliation{Department of Physics, Zhejiang University, Hangzhou, 310058, China}

\collaboration{(KMTNet Collaboration)}
\noaffiliation

\author{Martin Dominik}
\affiliation{Centre for Exoplanet Science, SUPA, School of Physics \& Astronomy, University of St Andrews, North Haugh, St Andrews KY16 9SS, UK}

\author{Christiane Helling}
\affiliation{Centre for Exoplanet Science, SUPA, School of Physics \& Astronomy, University of St Andrews, North Haugh, St Andrews KY16 9SS, UK}

\author{Markus Hundertmark}
\affiliation{Astronomisches Rechen-Institut, Zentrum f{\"u}r Astronomie der Universit{\"a}t Heidelberg (ZAH), 69120 Heidelberg, Germany}

\author{Uffe G. J{\o}rgensen}
\affiliation{Niels Bohr Institute \& Centre for Star and Planet Formation, University of Copenhagen, {\O}ster Voldgade 5, 1350 Copenhagen, Denmark}

\author{Penelope Longa-Pe{\~n}a}
\affiliation{Unidad de Astronom{\'{\i}}a, Facultad de Ciencias B{\'a}sicas, Universidad de Antofagasta, Av.\ Angamos 601, Antofagasta, Chile}

\author{Stephen Lowry}
\affiliation{Centre for Astrophysics \&\ Planetary Science, The University of Kent, Canterbury CT2 7NH, UK}

\author{Sedighe Sajadian}
\affiliation{Department of Physics, Isfahan University of Technology, Isfahan 84156-83111, Iran}

\author{Martin J. Burgdorf} 
\affiliation{Universit{\"a}t Hamburg, Faculty of Mathematics, Informatics and Natural Sciences, Department of Earth Sciences, Meteorological Institute, Bundesstra\ss{}e 55, 20146 Hamburg, Germany}

\author{Justyn Campbell-White} 
\affiliation{Centre for Astrophysics \&\ Planetary Science, The University of Kent, Canterbury CT2 7NH, UK}

\author{Simona Ciceri}
\affiliation{Department of Astronomy, Stockholm University, Alba Nova University Center, 106 91, Stockholm, Sweden}

\author{Daniel F. Evans}
\affiliation{Astrophysics Group, Keele University, Staffordshire, ST5 5BG, UK}

\author{Yuri I. Fujii} 
\affiliation{Niels Bohr Institute \& Centre for Star and Planet Formation, University of Copenhagen, {\O}ster Voldgade 5, 1350 Copenhagen, Denmark}
\affiliation{Institute for Advanced Research, Nagoya University, Furo-cho, Chikusa-ku, Nagoya, 464-8601, Japan}
\affiliation{Department of Physics, Nagoya University, Furo-cho, Chikusa-ku, Nagoya 464-8602, Japan}

\author{Tobias C. Hinse}
\affiliation{Korea Astronomy and Space Science Institute, Daejon 34055, Republic of Korea}

\author{Sohrab Rahvar}
\affiliation{Department of Physics, Sharif University of Technology, PO Box 11155-9161 Tehran, Iran}

\author{Markus Rabus} 
\affiliation{Centro de Astroingenier\'ia, Facultad de F\'isica, Pontificia Universidad Cat\'olica de Chile, Av. Vicu\~na Mackenna 4860, Macul 7820436, Santiago, Chile} 
\affiliation{Max Planck Institute for Astronomy, K{\"o}nigstuhl 17, 69117 Heidelberg, Germany}

\author{Jesper Skottfelt} 
\affiliation{School of Physical Sciences, Faculty of Science, Technology, Engineering and Mathematics, The Open University, Walton Hall, Milton Keynes, MK7 6AA, UK} 
\affiliation{Niels Bohr Institute \& Centre for Star and Planet Formation, University of Copenhagen, {\O}ster Voldgade 5, 1350 Copenhagen, Denmark}

\author{Colin Snodgrass}
\affiliation{School of Physical Sciences, Faculty of Science, Technology, Engineering and Mathematics, The Open University, Walton Hall, Milton Keynes, MK7 6AA, UK} 

\author{John Southworth}
\affiliation{Astrophysics Group, Keele University, Staffordshire, ST5 5BG, UK}

\collaboration{(MiNDSTEp Collaboration)}
\noaffiliation

\correspondingauthor{Yossi Shvartzvald}
\email{yossishv@gmail.com}

\begin{abstract}
The kinematics of isolated brown dwarfs in the Galaxy, beyond the solar neighborhood, is virtually unknown.
Microlensing has the potential to probe this hidden population, as it can measure both the mass and five of the six phase-space coordinates (all except the radial velocity) even of a dark isolated lens.
However, the measurements of both the microlens parallax and finite-source effects are needed in order to recover the full information.
Here, we combine {\it Spitzer} satellite parallax measurement with the ground-based light curve, which exhibits strong finite-source effects, of event OGLE-2017-BLG-0896.
We find two degenerate solutions for the lens (due to the known satellite-parallax degeneracy), which are consistent with each other except for their proper motion.
The lens is an isolated brown dwarf with a mass of either $18\pm1M_J$ or $20\pm1M_J$.
This is the lowest isolated-object mass measurement to date,
only $\sim$45\% more massive than the theoretical deuterium-fusion boundary at solar metallicity,
which is the common definition of a free-floating planet.
The brown dwarf is located at either $3.9\pm0.1$ kpc or $4.1\pm0.1$ kpc toward the Galactic bulge,  but with proper motion in the opposite direction of disk stars, with one solution suggesting it is 
moving within the Galactic plane.
While it is possibly a halo brown dwarf, it might also represent a different, unknown population.
\end{abstract}


\section{{Introduction}
\label{sec:intro}}

The census, including kinematics, of luminous stars has been
rapidly improving over the past decade and has just taken
a further quantum leap with the publication of the Gaia DR2 data
release \citep{Gaia.2018arXiv.A}.  In general, it is usually supposed that low-mass brown dwarfs, 
which are essentially invisible beyond the immediate solar neighborhood,
share the kinematics of ``normal'' stars.  While there are no theoretical
arguments against this hypothesis, neither is there any observational
evidence in its favor.

{\it Spitzer} microlensing offers a unique opportunity
to probe the kinematics of low-mass objects.  From 2014-2018,
{\it Spitzer} has been observing a total of nearly 1000 microlensing
events toward the Galactic bulge 
\citep{Gould.2013.prop.A,Gould.2014.prop.A,Gould.2015.prop.A,Gould.2015.prop.B,Gould.2016.prop.A} with the
aim of measuring their microlens parallax, $\bpi_\e$,
\begin{equation}
\bpi_\e = \frac{\pi_\rel}{\theta_\e}\,\frac{\bmu_\rel}{\mu_\rel};
\qquad
\theta_\e=\sqrt{\kappa M\pi_\rel};
\qquad
\kappa\equiv \frac{4G}{c^2\au}\simeq 8.14\,\frac{\mas}{M_\odot},
\label{eqn:bpie}
\end{equation}
where $(\pi_\rel,\bmu_\rel)$
are the lens-source relative (parallax, proper motion) and $M$
is the mass of the lens.
For special cases in which the angular Einstein radius $\theta_\e$ is
measured, the {\it Spitzer} measurement of $\bpi_\e$ then yields
$M$ and $(\pi_\rel,\bmu_\rel)$.
\begin{equation}
M_l = \frac{\theta_\e}{\kappa\pi_\e} ;
\qquad
\pi_\rel = \theta_\e\pi_\e ;
\quad
\bmu_\rel = \frac{\theta_\e}{t_\e}\,\frac{\bpi_\e}{\pi_\e},
\label{eqn:mpirel}
\end{equation}
where $t_\e$ is the Einstein timescale of the microlensing event.
Then, if the source parallax $\pi_s$ and proper motion $\bmu_s$
are independently measured, one can infer five of the six
phase-space coordinates of the lens (even if it is dark), i.e.,
its position on the sky and 
\begin{equation}
\pi_l = \pi_\rel + \pi_s;
\qquad
\bmu_l = \bmu_\rel + \bmu_s .
\label{eqn:pilbmul}
\end{equation}

The key additional step (assuming that $\bpi_\e$ is measured)
is to measure $\theta_\e$.  For luminous lenses, this can in principle
be done by waiting until the lens is well separated from the source,
when they can be separately imaged.  In this case, their observed 
separation $\Delta\btheta$  immediately gives
$\bmu_\rel=\Delta\btheta/\Delta t$, where $\Delta t$ is the elapsed
time since the event.  To date such measurements are relatively rare
\citep{Alcock.2001.A,Batista.2015.A,Bennett.2015.A}
because one must wait more than 10 years for typical events to separate, but
with next generation (``30m'') telescopes, they are likely to become routine.

However, for dark lenses, there are only two known methods to measure
$\theta_\e$: astrometric microlensing \citep{Miyamoto.1995.A,Hog.1995.A,Walker.1995.A}
and finite-source effects \citep{Gould.1994.B}.  Astrometric microlensing
is not generally well-suited to low-mass lenses because their
$\theta_\e$ are small\footnote{The angular Einstein radius of a 0.05$M_\odot$ BD at 4 kpc is $\theta_E$=0.23 mas.
Thus, its maximal astrometric shift is only $\delta\theta_c\approx0.35\theta_E\approx0.09$ mas.}.  Moreover, while it is
a potentially powerful approach for high-mass lenses (e.g., \citealt{Gould.2014.B}),
it can only be applied to a tiny handful of events with current 
instruments.  This implies that measuring finite-source effects
(together with microlens parallaxes) is presently the only viable
method to acquire a sample of low-mass dark lenses with measured
kinematics.

{\it Spitzer} microlensing is providing a steady stream of isolated-object
mass measurements that is strongly biased toward both low-mass lenses and
bright sources.  The latter enable relatively easy measurements of
$\bmu_s$, while $\pi_s$ is reasonably well known for essentially all
microlensing events.  With these quantities one can apply
Equations~(\ref{eqn:mpirel}) and (\ref{eqn:pilbmul}) to obtain 
the lens kinematics.

Finite-source effects (i.e., deviations in the light curve relative
to the predictions for a point source), occur when a source transits
a caustic in the magnification structure (or comes very close to a cusp).
This occurs relatively frequently for binary and planetary events because
the binary caustic structures are relatively large while the events
are usually recognized as planetary in nature because the source
passes over or very near a caustic.  However, for isolated lenses,
the ``caustic'' consists of a single point, i.e., directly behind the lens itself.
Thus, the probability of such a caustic passage (given that there is
a microlensing event) is
\begin{equation}
P = \rho\equiv \frac{\theta_*}{\theta_\e},
\label{eqn:rhodef}
\end{equation}
where $\theta_*$ is the source angular size.
This simple equation has two very important implications.  First,
it means that the rate $\Gamma_{\rm FS}=\rho\Gamma_{\mu\rm lens}$ of events with 
finite source effects does
not depend on the mass of some class of lenses, but only
on their number density $n$ \citep{Gould.2012.A}.  That is, while the microlensing
rate $\Gamma_{\mu\rm lens}\propto n\mu_\rel\theta_\e$ increases with mass as
$\Gamma_{\mu\rm lens}\propto M^{1/2}$, the finite-source
rate 
\begin{equation}
\Gamma_{\rm FS}=\rho\Gamma_{\mu\rm lens}\propto n\mu_\rel\theta_*
\label{eqn:gammafs}
\end{equation}
does not.  Thus, there is a strong bias toward the more common low-mass objects \citep{Kroupa.2001.A,Chabrier.2003.A}.
Second, because (from Equation~(\ref{eqn:gammafs})) 
$\Gamma_{\rm FS}\propto\theta_*$,
finite-source effects are strongly biased toward large (hence, bright)
stars.

There are four published isolated-object mass measurements from 
{\it Spitzer} microlensing in 2015 and 2016 \citep{Zhu.2016.A,Chung.2017.A,Shin.2018.arXiv.A},
and four more that we have identified from {\it Spitzer} microlensing
in 2017.  These have masses in ascending order,
$M=(
19,  
45,  
58^\dagger,  
80,  
88,  
235^\dagger, 
520, 
570^\dagger  
)M_{\rm jup}$,
which illustrates the strong bias toward low mass objects.  Here the
``$\dagger$'' symbol indicates preliminary estimates for not-yet-published
events.
Their source radii are (re-sorted in ascending order)
$\theta_*=(
1.4,  
5.7,  
5.8,  
6.0^\dagger,  
6.3, 
6.8^\dagger, 
7.8,  
33.7^\dagger  
)\muas$, 
which should be compared to $\theta_*\sim 0.5\muas$ for typical
microlensing events.

Here we present the first of the 2017 {\it Spitzer} 
isolated-object microlensing mass measurements, OGLE-2017-BLG-0896L.  
As we will report, it has $M\simeq19\,M_J$, making it the lowest-mass
object of the sample of eight that have been measured to date.  Indeed,
this was the initial focus of our interest.  However, in the course
of checking our results, we noted that the values of 
$(\pi_\rel,\bmu_\rel)$, which are automatically returned as part of
the mass derivation, pointed to a possible conflict with the known
kinematic characteristics of the major populations of the Galaxy.
Because this discrepancy could be resolved if the source had
mildly unusual characteristics, we undertook the additional step
of measuring the source proper motion $\bmu_s$.  Contrary to
our expectation, this measurement made the conflict substantially
worse.  Of course, one cannot draw very strong conclusions from a single
unusual object.  However, as we note, there are at least some indications
that this object may be a member of previously unrecognized population
We discuss this possibility, as well as possible biases of the $Spitzer$ program favoring the detections of such objects, in Section~\ref{sec:discuss}.

\section{{Observations}
\label{sec:obs}}

OGLE-2017-BLG-0896 is at (RA,Dec)$_{J2000}$ = (17:39:30.98,$-27$:17:51.1)
corresponding to $(l,b)=(0.69\degree,2.01\degree)$.  
It was discovered and announced as a probable microlensing event
by the OGLE Early Warning
System \citep{Udalski.1994.A,Udalski.2003.A} at UT 20:23 on 25 May 2017.
The event lies
in OGLE field BLG675 \citep{Udalski.2015.B}, for which OGLE observations
were at a cadence of 1--3 obs/night using their 1.3m telescope
at Las Campanas, Chile.

The Korea Microlensing Telescope Network (KMTNet, \citealt{Kim.2016.A})
observed this field from its three 1.6m
telescopes at CTIO (Chile, KMTC), SAAO (South Africa, KMTS) and SSO 
(Australia, KMTA),
in its field BLG15 with a cadence of 1 obs/hr.  
It is designated SAO15N0405.007056 in the KMTNet catalog.
We exclude for the fit the KMTNet data over the peak of the event, from HJD'$\equiv$ HJD$-2450000$=7910 to HJD'=7912,
as the event got too bright and thus the photometry is affected by nonlinearity. 

The great majority of these survey observations were carried out in 
the $I$ band 
with occasional $V$-band observations made
solely to determine source colors.
All reductions for the light curve
analysis were conducted using variants of difference image analysis (DIA,
\citealt{Alard.1998.A}), specifically \citet{Wozniak.2000.A} and \citet{Albrow.2009.A}.

OGLE-2017-BLG-0896 was announced as a {\it Spitzer} target at UT 09:21
on 5 June 2017 because it was recognized as a relatively
high-magnification event $A_\max\ga 20$ and so with good \citep{Gould.1992.A,Abe.2013.A}
or possibly excellent \citep{Griest.1998.A} sensitivity to planets.
The {\it Spitzer} observations themselves could not begin until 17 days
later, when the event entered the sun-angle window, which was
coincidentally the first epoch of planned observations, beginning
UT 15:46 June 22, 2017.
The $Spitzer$ data were reduced using the \cite{Calchi.2015.B} algorithm for crowded-field photometry.

The {\it Spitzer} team alerted the event as high-magnification and mobilized intensive follow-up
observations, with the aim of detecting and characterizing any planetary signatures.
Follow-up observations were carried out using four of the Las Cumbres Observatory (LCO) 
global network of telescopes in Chile, South Africa, and Australia, with the SDSS-$i'$ filter.
The Microlensing Follow Up Network ($\mu$FUN) followed the event using
the 1.3m SMARTS telescope at CTIO (CT13) with $V/I/H$-bands, the 0.4m telescope at Auckland Observatory (AO) with $R$-band,
the 0.3m Perth Exoplanet Survey Telescope (PEST) at Perth, Western Australia,
and the 0.25m telescope at Craigie, Western Australia (unfiltered).
PEST data were excluded from the analysis due to systematics of unknown origin.
The MiNDSTEp team followed the event using the Danish 1.54-m telescope hosted at ESO's La Silla observatory in Chile,
with a simultaneous two-color instrument (wide visible and red; See Figure~1 of \citealt{Evans.2016.A}) providing Lucky Imaging photometry (\citealt{Skottfelt.2015.A}).
For the analysis of the event we use only the Danish red-band data.
LCO and AO data were reduced using pySIS \citep{Albrow.2009.A}, CT13 and Craigie data were reduced using DoPhot \citep{Schechter.1993.A},
and Danish data were reduced using  a modified version of DanDIA \citep{Bramich.2008.A}.

\begin{figure}[ht]
\plotone{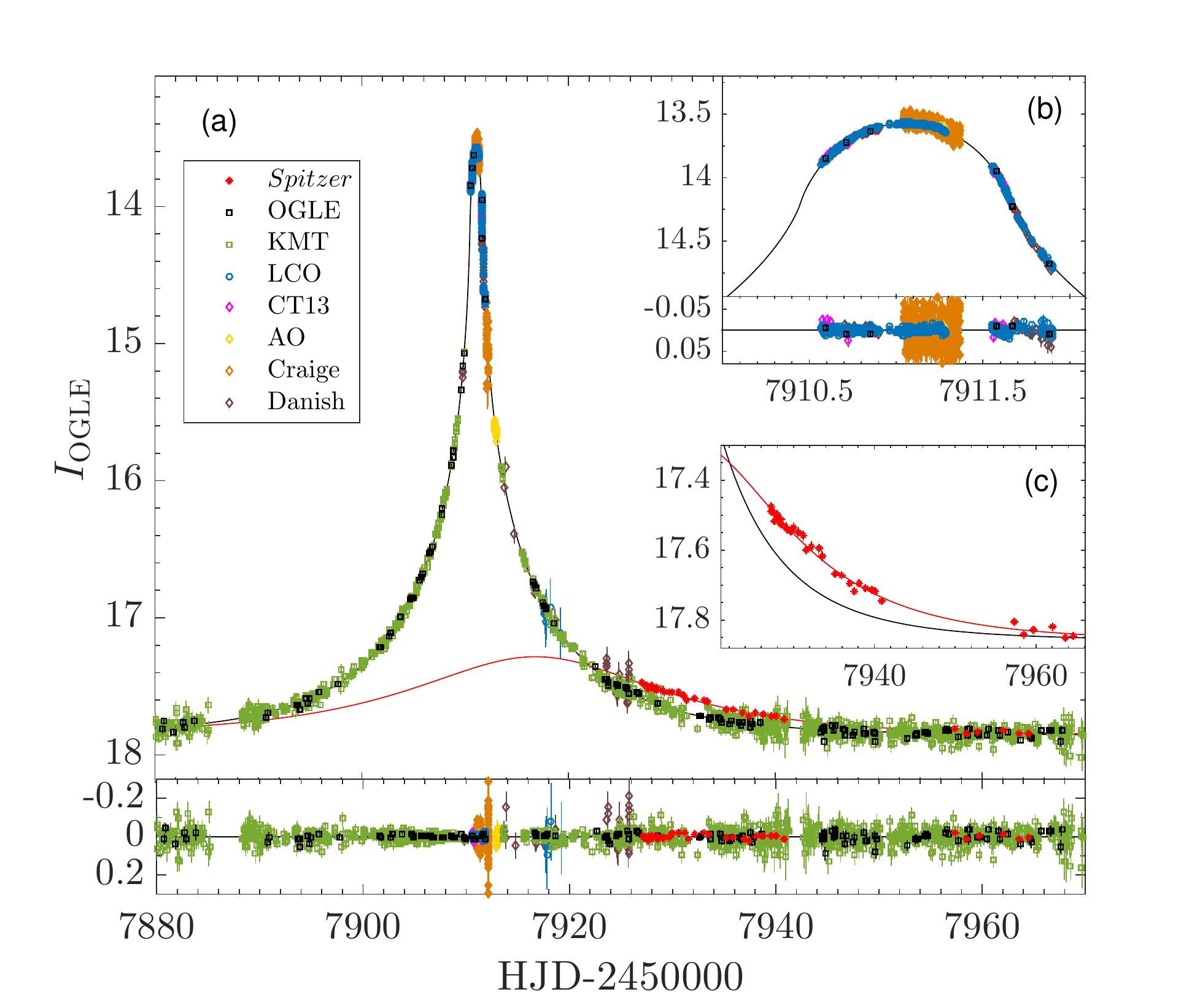}
\caption{Light curve of OGLE-2017-BLG-0896. Finite-source effects are clearly seen at the peak of the event (inset b).
The $Spitzer$ light curve is significantly offset from the ground-based model (inset c), indicating the large microlens parallax.
}
\label{fig:lc}
\end{figure}

While no planetary anomalies were detected, the follow-up observations were crucial in order to model the finite-source effects that are clearly shown at the peak of the event (see Figure~\ref{fig:lc}) because the KMTNet data over the peak were affected by nonlinearity and OGLE cadence was not sufficient for the characterization.

\section{{Light Curve Analysis}
\label{sec:anal}}

\subsection{Ground data only}

The light curve, as seen from Earth, is of a symmetric high-magnification event with clear deviation from a point source microlensing (Figure~\ref{fig:lc}).
These features rule out any reasonable binary lens since no anomaly/asymmetry associated with a central caustic is detected (see Section~\ref{sec:dist_host}).  
The $Spitzer$ data cover only the falling tail of the event, thus not constraining the finite-source size. Therefore, we start by modeling the ground-based data alone.

We fit the ground-based light curve using 
six parameters to describe the geometry of finite-source point-lens
(FSPL) microlensing as well as
two flux parameters for each dataset, $f_{s,i}, f_{b,i}$ (for the source and possible blend).
The geometric parameters are the Paczy{\'n}ski parameters, $(t_0,u_0,t_\e)$ \citep{Paczynski.1986.A}, the scaled angular source size $\rho= \theta_*/\theta_{\rm E}$, and the limb-darkening coefficients $\Gamma_I$ and $\Gamma_{\rm Danish}$ (we use a specific coefficient for the Danish data because of the non-standard filter).
We adopt a limb-darkened brightness profile for the source star of the form
\begin{equation}
S_{\lambda}(\phi) = \bar{S}_{\lambda}\left[1 - \Gamma_{\lambda}\left(1 - \frac{3}{2}\cos\phi\right)\right],
\end{equation}
where $\bar{S}_{\lambda} \equiv f_{s, \lambda}/(\pi \theta^{2}_{\star})$ is the mean surface brightness of the source,  $\phi$ is the angle between the normal to the surface of the source star and the line of sight, $f_{s,\lambda}$ is the total source flux and $\Gamma_{\lambda}$ is the limb-darkening coefficient at wavelength $\lambda$, respectively \citep{An.2002.A}.
The limb-darkening coefficients are usually estimated using the source intrinsic properties, which are interpreted from the offset between its observed color and magnitude and the red clump centroid. For this interpretation one assumes that the source is at a similar distance as the clump (i.e., in the bulge).
In the case of OGLE-2017-BLG-0896L, the dense coverage during the finite-source effects allows us 
to well constrain the limb-darkening coefficient, $\Gamma_I$, thus enabling us to verify that indeed the source is a bulge star (see Section~\ref{sec:source}). We use $\Gamma_I$ and $\Gamma_{\rm Danish}$ as free fit parameters, 
as most of our observations over the peak are with these bands\footnote{We use $\Gamma_I$ also for LCO SDSS-$i'$ data.}.
For AO ($R$-band) and Craigie (unfiltered) data, we estimate the limb-darkening coefficient as $(\Gamma_I+\Gamma_V)/2$, where $\Gamma_V=0.754$ was determined from \cite{Claret.2011.A} based on the characterized source properties (Section~\ref{sec:source}). The $V$- (OGLE/KMTNet/CT13) and $H$-band (CT13) data are used only to derive the source color, and thus do not require limb-darkening coefficients.

\subsection{Satellite parallax degeneracy}

In order to include the $Spitzer$ data we add two microlensing parallax parameters, $\pi_{\rm E,N}, \pi_{\rm E,E}$, aligned with the equatorial north and east directions. Generally, this can introduce the well known four-fold satellite parallax degeneracy \citep{Refsdal.1966.A,Gould.1994.A}. However, because $|u_{0,ground}|\ll 1$ 
the magnitude $|\bpi_\e|$ is nearly the same for all solutions \citep{Gould.2012.A}, and thus the mass and distance of the lensing system are similar.
A two-fold degeneracy in the direction of the relative proper motion between the source and the lens persists.

Because $Spitzer$ data covered only the falling part of the event and in addition did not fully cover the baseline of the event (see inset of Figure~\ref{fig:lc}), they cannot set strong constraints on $\bpi_\e$ by themselves. However, by applying a constraint on the $Spitzer$ source flux based on color-color relations, the parallax measurement can be significantly improved (e.g., \citealt{Calchi.2015.A}). 
We derive two color-color relations for OGLE-2017-BLG-0896:
a $VIL$ relation (using KMTNet data) and an $IHL$ relation (using CT13 data), as detailed in Section~\ref{sec:cmd}.
The constraints on $Spitzer$ source flux using each of the relations, and consequently the derived parallax values, are in good agreement with each other ($<1\sigma$).
We adopt the $VIL$ relation for the final results, because the CT13 data might be subject to low-level chromatic effects.

Table~\ref{tab:model} gives the best-fit parameters and their uncertainties for the four-fold degenerate solutions ($\Delta\chi^2<4$), which were found using ``Newton's method'' (\citealt{Simpson.1740.A}; see \citealt{Skowron.2012.arXiv.A}).
The microlensing parallax components are well constrained, with $\sim4\%$ and $\sim8\%$ uncertainties on $\pi_{\rm E,N}$ and  $\pi_{\rm E,E}$, respectively.
These are significantly better than the results without the constraint on $Spitzer$ flux, which have $15-30\%$ uncertainties on the parallax components.
It is important to note, however, that the median values are similar. In particular, $\pi_{\rm E,E}<0$ at the 3$\sigma$ level even without the color constraint, which is both surprising and interesting as we discuss below in Section~\ref{sec:lens}.

\subsubsection{{Negative $Spitzer$ blending}
\label{sec:flux}}

The $Spitzer$ instrumental blend flux is constrained to be negative when using the color-color relations, $f_{b,Spitzer} = -4.4\pm1.2$.
While negative blending is known to sometimes be present in ground-based microlensing light curves (e.g., \citealt{Jiang.2004.A}),
its origin in these cases is not always clear. However, for $Spitzer$ photometry in crowded fields using the \cite{Calchi.2015.B} algorithm, the cause for possible artificial negative blending is well understood. As detailed in \cite{Calchi.2015.B}, an input catalog of sources is used to retrieve the $Spitzer$ photometry around the event.
The catalog is constructed from optical survey data (KMTNet data in the case of OGLE-2017-BLG-0896), which have better resolution and depth than the $Spitzer$ image.
Any source that is not in the catalog (i.e., unresolved faint stars) will be absorbed in the global background flux, which effectively is subtracted from the source flux, thus resulting as an artificial negative blending. Naturally, this will be more significant in cases for which no real underlying blend in the source position is detected, like in our case ($f_{b,ogle} = 0.028\pm0.009$, corresponding to 5$\sigma$ limit of $I_b>20.8$).

Examining the optical image around the event and comparing it to nearby ($<15''$) isolated regions, we find an excess of flux due to unresolved stars. The $Spitzer$ flux in the isolated regions is significantly lower than the background estimation at the source position. After taking into account $Spitzer$ point-response function, this difference correspond to $\sim$5 flux units of artificial negative blending, which therefore fully explains the negative blend found for $Spitzer$.

\section{{Source star}
\label{sec:source}}

\subsection{{CMD analysis and color-color relations}
\label{sec:cmd}}

The source photometric properties (color and magnitude) are important for several reasons.
First, the source intrinsic properties yield its angular size, $\theta_*$, which is needed to derive $\theta_{\rm E}$ and the physical properties of the lensing system (Equation~(\ref{eqn:mpirel})). Second, they are used to estimate the limb-darkening coefficients, or alternatively (as in our case) can be compared to the fitted coefficients to verify the estimate of the distance to the source. Lastly, instrumental color-color relations can help constrain the source flux in a third band based on one measured color (e.g., the $Spitzer$ $L$-band source flux based on an optical color).

\begin{figure}[ht]
\plotone{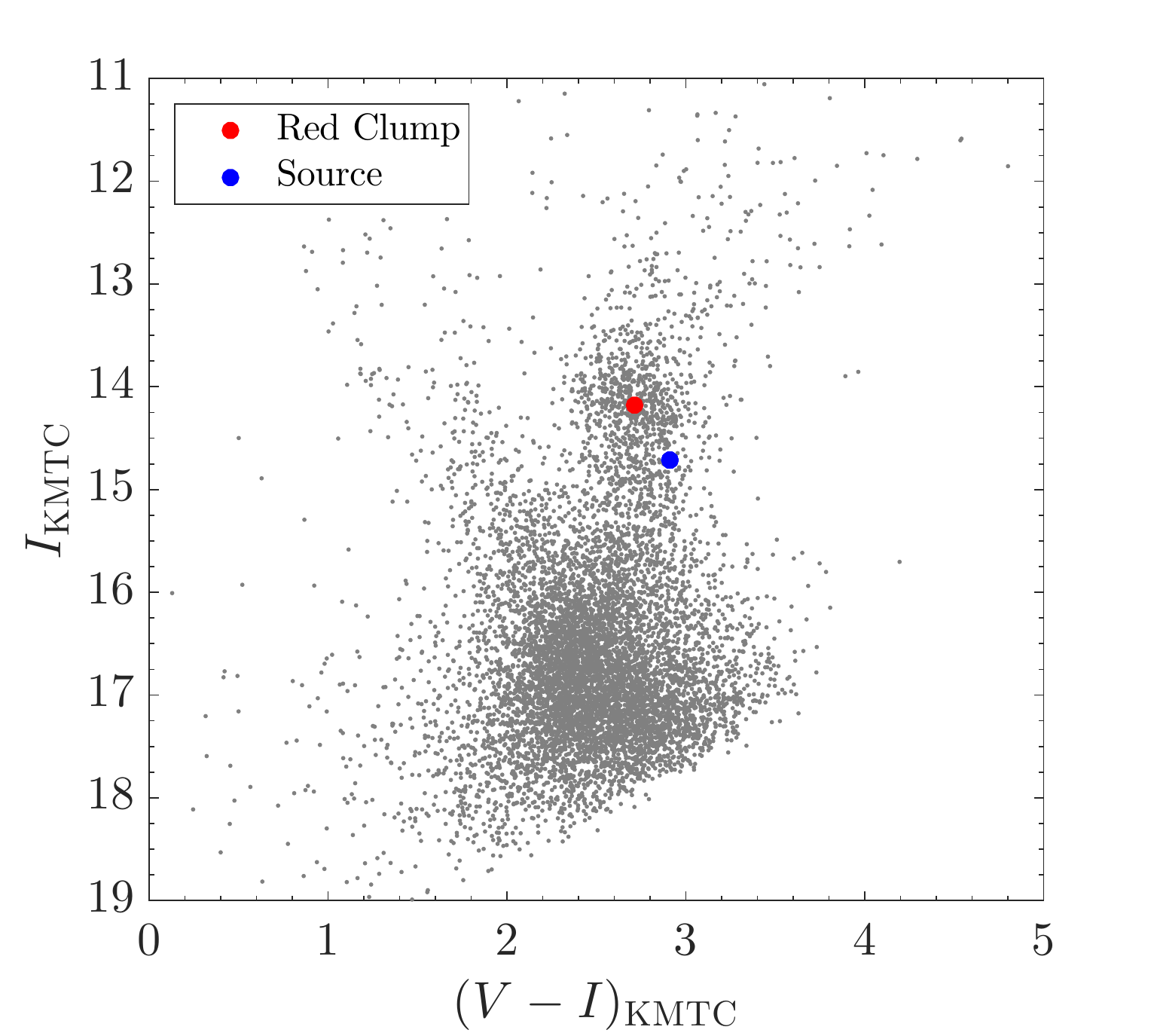}
\caption{KMTC instrumental color-magnitude diagram of OGLE-2017-BLG-0896. The source angular size $\theta_*$ is derived using the
offset between the red clump (red circle) and the source (blue circle) positions.
}
\label{fig:cmd}
\end{figure}

Figure~\ref{fig:cmd} shows the KMTC $V/I$ instrumental color-magnitude diagram (CMD) constructed from sources within $<2'$ of the event.
We use the method described in \cite{Nataf.2013.A} to measure the instrumental centroid of the red clump 
$(V-I,I)_{\rm cl, kmt}=(2.71 ,14.18)$ 
and compare it to the intrinsic centroid of $(V-I,I)_{\rm cl,0}=(1.06,14.41)$ 
\citep{Bensby.2013.A,Nataf.2013.A}. We determine the instrumental source color from
regression of $V$ versus $I$ flux as the source magnification changes
\citep{Gould.2010.A}, and find $(V-I)_{s, {\rm  kmt}}=2.91\pm0.03$.
The source instrumental magnitude, as inferred from the microlensing model, is 
$I_{s, {\rm kmt}}=14.72\pm 0.01$. 
Assuming that the source lies behind the same dust
column as the red clump, its intrinsic properties are $(V-I,I)_{s,0}=(1.26,14.95)\pm(0.06, 0.04)$, accounting also for the red clump instrumental and intrinsic uncertainties.
Using standard color-color relations \citep{Bessell.1988.A} and
the relation between angular source size and surface brightness
\citep{Kervella.2004.A}, we find $\theta_*=5.71\pm 0.29\,\muas$.

The source position on the CMD, under the assumption it is a bulge star, suggests a K2.5 III spectral type with $T_{\rm eff}\approx4300$ and log($g$)$\approx2.2$.
The corresponding linear limb-darkening coefficients \citep{Claret.2011.A} are $\Gamma_I=0.519\pm0.015$ and $\Gamma_V=0.754\pm0.021$, where the uncertainties account for a range of possible metallicities and microturbulence velocities.
The limb-darkening coefficient $\Gamma_I$ derived from the fit, for all four degenerate solutions (see Table~\ref{tab:model}), is within excellent agreement of the estimate based on the source spectral type. This confirms the assumption of a bulge source with similar distance as the red clump.
We note that the derived $\Gamma_I$ can also explain M/K dwarfs. However, these would be either significantly fainter (if in the bulge) or significantly bluer (if nearby).

We extract $Spitzer$ photometry for red giant branch stars 
($13.7<I_{\rm KMT}<14.7\,;\,2.45<(V-I)_{\rm KMT}<2.95$), 
which are a good representation of the bulge giant population, 
and derive an instrumental $VIL$ color-color relation \citep{Calchi.2015.B}.
Applying the relation to the measured $(V-I)_{s,\rm KMT}$,
we find $(I_{\rm KMT}-L_{Spitzer})_s=0.31\pm 0.05$.
Using this constraint in the microlensing model gives $f_{s,Spitzer}=27.2\pm1.2$.
For consistency, we also derive the instrumental $IHL$ relation using CT13 data.
Applying it to $(I-H)_{s,\rm CT13}=0.53\pm0.02$ (derived from regression),
we find $(I_{\rm CT13}-L_{Spitzer})_s=4.45\pm 0.03$.
This gives $f_{s,Spitzer}=27.0\pm0.7$, in excellent agreement with the constraint using the $VIL$ relation.
We note that almost all CT13 data (except 3 baseline epochs) were taken during the finite-source effects, and thus they might exhibit  
low-level chromatic effects.

\subsection{{Source proper motion}
\label{sec:pms}}

\begin{figure}[ht]
   \centering
  \includegraphics[width=0.7\textwidth]{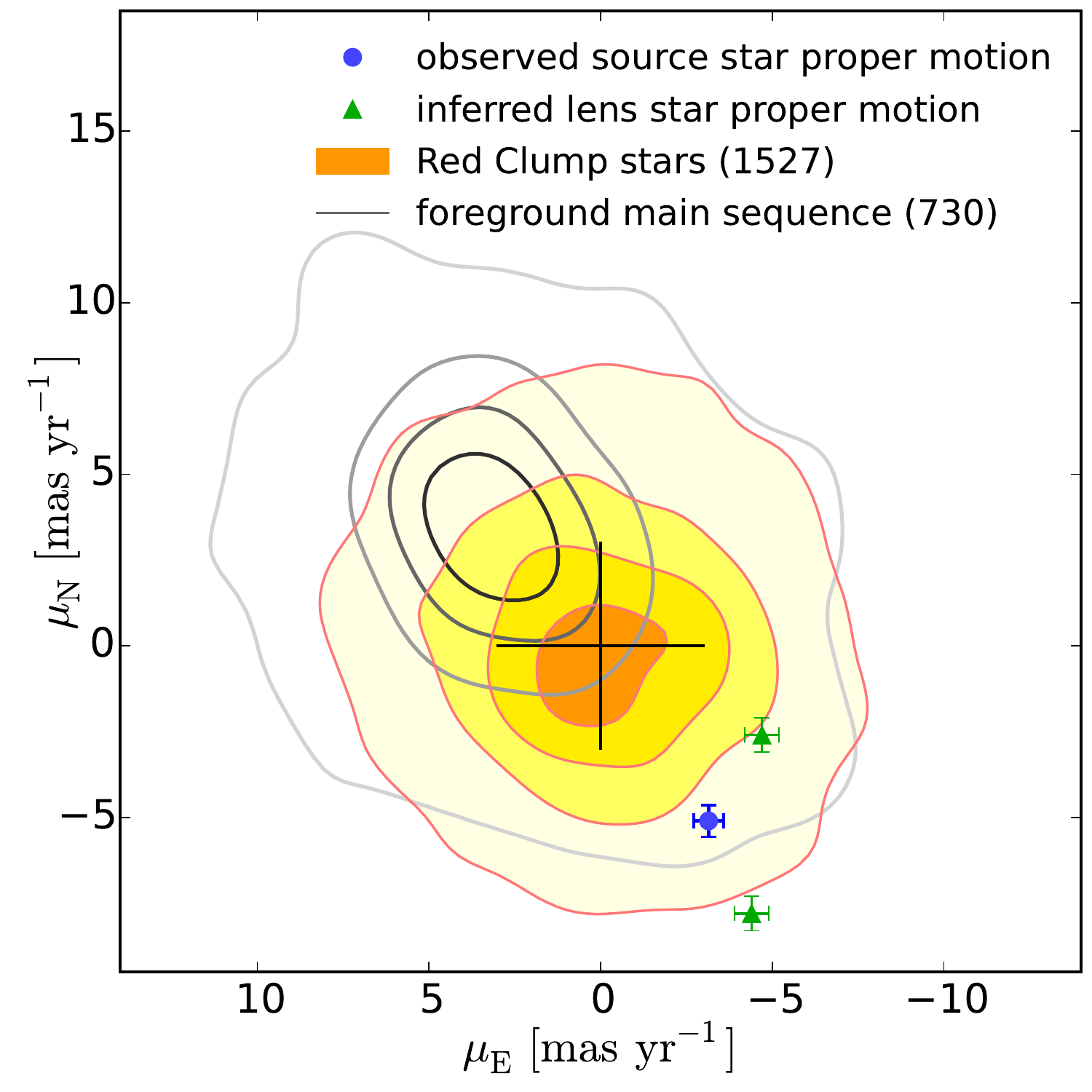}
\caption{OGLE proper motion of bulge (red clump stars) and disk (foreground stars) populations.
The contours contain 95.34\%, 68.36\%, 45.89\%, 24.11\% of the stars in each population.
We note that the foreground star population represents stars at various distances and thus an effective integration over a range of proper-motion distributions, where nearby stars have the largest proper-motion dispersion.
The source observed proper motion (blue circle) and the two degenerate solutions for the lens proper motion (green triangles) are shown.
 \label{fig:pm}}  
\end{figure}

The lens proper motion can be derived from the relative-proper motion and the source proper motion (Equation~(\ref{eqn:pilbmul})).
The source star of OGLE-2017-BLG-0896 is bright, isolated and with negligible blending (the blend is at least 3 magniutdes fainter), thus permitting a good measurement of its proper motion (unlike most microlensing sources, which are faint and blended).
We construct a deep OGLE CMD from a $6.5'\times6.5'$ region centered around the event, and identify 1527 red clump bulge stars and 730 foreground disk stars.
We then use 387 good seeing ($<$1.35'') OGLE epochs from HJD'=5385--8030 to measure the vector proper motion of each star, with typical uncertainty of 0.45 $\masyr$ for clump stars.
We find that the source proper motion is
$\mu_s(N,E) = (-5.10,-3.15)\pm(0.46,0.44) \masyr$ relative to the frame set by the clump giants.
Figure~\ref{fig:pm} shows the source proper motion along with the proper motion distributions of bulge and disk stars.
The position of the source on this diagram further supports it being part of the bulge population. 

\section{{The Lens - Counter-Rotating Brown Dwarf}
\label{sec:lens}}

The Einstein angular radius is determined by combining $\rho$ from the model and $\theta_*$ from the CMD,
\begin{equation}
\theta_{\rm E} = 0.140\pm0.007\,\mas.
\end{equation}
Combining this with the four-degenerate parallax solutions from the microlensing model (Equation~(\ref{eqn:mpirel})) yields a low-mass BD of $M\simeq19\, M_J$, with minor differences within 1-2$\sigma$ between the models (See Table~\ref{tab:phys}).
The distance to the BD (Equation~(\ref{eqn:pilbmul})) is $D_l\simeq4$ kpc, where we assumed $D_S=8.3$ kpc, which is appropriate for a bulge source toward the event direction.

The geocentric relative proper motion (Equation~(\ref{eqn:mpirel})) is $\mu_{\rm rel, geo}=3.42\pm0.18\, \masyr$, with either a North-West or a South-West direction as inferred from the parallax components. These already suggest some tension with a disk lens (as would seem to be inferred by $D_l$). In principle, this tension could be resolved if the source had significant North-East proper motion. However, as we found in Section~\ref{sec:pms}, the source is actually moving in the opposite direction.
Accounting for Earth's projected velocity at the peak of the event, $V_{\oplus,\perp}(N,E)=(-0.68,28.9)\,\kms$, the two degenerate solutions for the BD heliocentric proper motion relative to the frame set by the bulge clump giants are (see Figure~\ref{fig:pm}),
\begin{equation}
\begin{aligned}
\mu_{l,{\rm hel}}(N,E) = {\bmu_s}+({\bmu_{\rm rel,\geo}}+\pi_{\rm rel}{\bdv V}_{\oplus,\perp})=\left\{
  \begin{array}{c}
    (-7.8,-4.4)\pm(0.5,0.5)\,\masyr\\
    {\rm or}\\
    (-2.5,-4.7)\pm(0.5,0.5)\,\masyr
  \end{array}
\right.
\end{aligned}
\label{eqn:mul}
\end{equation}

In order to find the lens projected velocity, we first note that
\begin{equation}
\begin{aligned}
 \bmu_{l,{\rm hel}} = {\bmu_s}+{\bmu_{\rm rel, hel}}=
 \left(\frac{{\bdv V}_s-{\bdv V}_{\odot}}{D_s}-\frac{{\bdv V}_{\rm cl}-{\bdv V}_{\odot}}{D_{\rm cl}}\right)+
 \left(\frac{{\bdv V}_l-{\bdv V}_{\odot}}{D_l}-\frac{{\bdv V}_s-{\bdv V}_{\odot}}{D_s}\right)
 \end{aligned}
\end{equation}
where ${\bdv V}_{\rm cl}$ and $D_{\rm cl}$ are, respectively, the mean velocity and distance of the clump stars that set the proper-motion reference frame.
Because the event is at $l\approx0$, we adopt $V_{\rm cl}(l,b)=(0,0)\,\kms$ and $D_{\rm cl}=8.3$ kpc.
The Sun's velocity consists of peculiar velocity, $V_{\odot,\rm pec}(l,b) = (12,7)\,\kms$ \citep{Schonrich.2010.A}, and the disk circular velocity, $V_{\rm rot}(l,b) = (220,0)\,\kms$ \citep{Camarillo.2018.arXiv.A}. 
Therefore, the two degenerate solutions for the lens peculiar velocity relative to the mean motion disk stars in its neighborhood are
\begin{equation}
\begin{aligned}
V_{l,{\rm pec}}(l,b) = {\bdv V}_l-{\bdv V}_{\rm rot}=D_l{\bmu_{l,{\rm hel}}} - {\bdv V}_{\rm rot}\left( \frac{D_l}{D_{\rm cl}}\right)+ 
{\bdv V}_{\odot,\rm pec}\left( 1-\frac{D_l}{D_{\rm cl}}\right)\\
=\left\{
  \begin{array}{c}
    (-260,-3)\pm(10,9)\,\kms\\
    {\rm or}\\
    (-193,54)\pm(10,10)\,\kms
  \end{array}
\right.\quad\quad\quad\quad\quad\,\,\,
\end{aligned}
\label{eqn:vl}
\end{equation}
These should be compared to the standard deviations for the disk velocities of $\sigma_{\rm rot}(l,b) = (30,20)\,\kms$.
Thus, in both cases the BD is significantly counter-rotating relative to the disk-stars' motion.
Interestingly, one of these solutions is consistent, within small
error bars, with perfectly planar counter-rotation.  The other
solution has considerable out-of-plane motion.

\subsection{{Constraints on possible distant companion}
\label{sec:dist_host}}
Our data can rule out a distant companion to the BD via two channels. First, the flux from the companion cannot exceed the limits we found on the blend flux in Section~\ref{sec:flux} ($I_b>20.8$). We conservatively assume that the lensing system suffers the same extinction as the red clump ($A_I\approx2.7$) and 
use PARSEC-COLIBRI isochrones \citep{Marigo.2017.A} to calculate the brightness of possible companions at the distance of the BD.
We find that the blend flux limit can exclude all main-sequence companions with $M>0.95M_\odot$.

The second constraint comes from the lack of additional features in the light curve.
These features can be either anomalies in the apparent single-lens light curve (e.g., features due to caustics) or an additional point-lens-like event if the the source passes near the companion (for more details see \citealt{Han.2005.A}). We follow the procedures of \cite{Mroz.2018.A} and set a lower limit on the distance of a possible companion. In short, we simulated binary-lens light curves with a companion at a range of separations, with a range of masses and at all possible projected angles. We calculate the fraction of light curves that show additional features (using a threshold of $\chi^2_{\rm single}-\chi^2_{\rm binary}>60$) and consider a detection if 90\% of the light curves pass this threshold. We find that companions with $M=0.95M_\odot$ (the upper limit from lens flux) can be excluded for separations $a_\perp<65$ AU, and companions with $M=0.08M_\odot$ can be excluded for separations $a_\perp<20$ AU.

The remaining parameter space of possible luminous companions (i.e., M/K dwarfs at separations $a_\perp>20$ AU) can be explored using future AO imaging, searching for any light from the putative companion \citep{Gould.2016.A}. This study can be done at first light of next generation (``30 meter'') telescopes (perhaps 2028), as the lensing system will be separated by more than 50 mas from the source by then, and thus clearly resolved if luminous.

\section{{Discussion}
\label{sec:discuss}}

We have presented the discovery of an $\sim$19$M_J$ isolated BD, the lowest-mass isolated object ever measured.
The BD is located at $D_l\simeq4$ kpc toward the Galactic bulge, but it is counter-rotating with respect to the kinematics of ``normal'' disk stars at this location.
This is not the first example of a low-mass object with unusual kinematics. OGLE-2016-BLG-1195L \citep{Shvartzvald.2017.B} is a planetary system at $D_l\simeq4$ kpc with an Earth-mass planet orbiting an ultracool dwarf ($\sim$0.08$M_\odot$), with significant West relative proper motion, $\mu_{\rm hel, E}\simeq -7.5\,\masyr$, although in that case the source proper motion was not measured and thus a fast moving source ($\sim 360\,\kms$ relative to the bulge) is also possible.
OGLE-2016-BLG-0864L (Chung et al. 2018, in preparation) is a BD-BD binary system at $D_l\simeq3$ kpc, with relative proper motion suggesting the system is counter-rotating with respect to the disk motion (though, again, the source proper motion was not measured).
In addition, while most local BDs have similar kinematics as stars (e.g., \citealt{Faherty.2009.A}), there is a growing sample of local BDs \citep{Zhang.2017.A} associated with kinematics of halo stars, including even a counter-rotating BD \citep{Cushing.2009.A}.

The combined measurements of the satellite microlens parallax with $Spitzer$ and the detection of finite-source effects, enabled the full characterization of the BD properties accessible to microlensing (mass and five out of six phase-space coordinates).
Microlensing is the only technique that can characterize the kinematics of low-mass dark objects throughout the Galaxy. This method can also be extended to free-floating planets \citep{Henderson.2016.A,Gould.2016.A}.
A possible explanation of the kinematics of OGLE-2017-BLG-0896L is that it is a halo BD. 
Alternatively, it might suggest, along with the other examples mentioned above, the existence of a counter-rotating population of low-mass objects.
Counter-rotating stellar disk populations have been detected in other galaxies (e.g., \citealt{Rubin.1994.A,Pizzella.2014.A,Armstrong.2018.A}), suggesting  an occurrence rate of $\sim10\%$ for spirals and $\sim30\%$ for S0 galaxies \citep{Pizzella.2004.A}. The scale of the counter-rotating component can range from a few tens pc (e.g., \citealt{Corsini.2003.A}) to a few kpc (e.g., \citealt{Rubin.1994.A}).
While locally there is no evidence for a large counter-rotating population in our Galaxy, it may exist in the inner Galaxy.

The selection criteria of $Spitzer$ events \citep{Yee.2015.A}, with  the 3-10 day lag before event selection and the beginning of $Spitzer$ observations, is favoring the detection of these BDs, which have longer timescales than expected by ``normal'' disk star kinematics (e.g. OGLE-2017-BLG-0896L, OGLE-2016-BLG-1195L). 
In addition, counter-rotating lenses will peak later as seen from $Spitzer$ than from Earth, thus increasing the chances for parallax measurement.
This can be considered as a microlens-parallax ``Malmquist bias'', because events that will peak earlier for $Spitzer$ might already be at baseline by the time of first $Spitzer$ observations and thus the parallax will not be measured. The bias is mostly relevant for short $t_E$ events and faint high-magnification events. However, for events with typical timescale and peak magnification this bias should be small.

\acknowledgments 
We thank D. Kirkpatrick for fruitful discussions about BDs.
Work by YKJ, and AG were supported by AST-1516842 from the US NSF.
IGS, and AG were supported by JPL grant 1500811.
Work  by  C. Han was supported by the grant (2017R1A4A1015178) of
National Research Foundation of Korea.
This work is based (in part) on observations made with the $Spitzer$ Space
Telescope, which is operated by the Jet Propulsion Laboratory, California
Institute of Technology under a contract with NASA. Support for this work
was provided by NASA through an award issued by JPL/Caltech.
This work was partially supported by NASA contract NNG16PJ32C.
The OGLE project has received funding from the National Science Centre,
Poland, grant MAESTRO 2014/14/A/ST9/00121 to AU.
This research has made use of the KMTNet system operated by the Korea
Astronomy and Space Science Institute (KASI) and the data were obtained at
three host sites of CTIO in Chile, SAAO in South Africa, and SSO in
Australia.
Work by SR and SS was supported by INSF-95843339.
PL-P was supported by MINEDUC-UA project, code ANT 1656.
The research has made use of data obtained at the Danish 1.54m telescope at ESO’s La Silla Observatory.

\newpage

\begin{table}[ht]
\centering
\caption{ Microlensing model \label{tab:model}}
\begin{tabular}{|l|c|c|c|c|}
\tableline
         & $++$ & $+-$ & $-+$ & $--$\\
\tableline
 $\chi^2$    & 4126.8 & 4130.1 & 4127.0 & 4130.1\\
 \tableline
 $t_0$ [HJD']& 7911.05582(68) &7911.05601(68) & 7911.05578(68) & 7911.05601(68)\\
 \tableline
 $u_0$       & 0.0039(11) &0.0037(12) & -0.0038(11) & -0.0037(12)\\
 \tableline
 $t_{\rm E}$ [d] & 14.883(93) & 14.896(93)& 14.885(93) & 14.896(93)\\
 \tableline
 $\rho$      & 0.04092(31) & 0.04085(30)& 0.04091(30) & 0.04085(31)\\
 \tableline
 $\Gamma_I$  & 0.525(13) & 0.520(13)& 0.523(13) & 0.522(13)\\
 \tableline
 $\Gamma_{\rm Danish}$  & 0.454(23) & 0.450(23)& 0.453(23) & 0.450(23)\\
 \tableline
 $\pi_{EN}$  & -0.779(28) &0.662(29) &-0.771(28) & 0.669(29)\\
 \tableline
 $\pi_{EE}$  & -0.615(46) &-0.587(46) & -0.613(46) & -0.589(46)\\
 \tableline
\end{tabular}
\end{table}

\begin{table}[ht]
\centering
\caption{Physical properties \label{tab:phys}}
\begin{tabular}{|l|c|c|c|c|}
\tableline
         & $++$ & $+-$ & $-+$ & $--$\\
\tableline
$\theta_{\rm E}$ [mas]  & 0.1395(72) & 0.1398(72)& 0.1396(72)& 0.1398(72) \\
\tableline
$M_l$ [$M_J$]            & 18.1(1.0) & 20.3(1.2) & 18.2(1.0)& 20.2(1.1)\\
\tableline
$D_l$ [kpc]           & 3.86(11) & 4.10(12) & 3.88(11)& 4.08(12)\\
\tableline
$\mu_{\rm rel, geo}$ [$\masyr$]  & 3.42(18) & 3.43(18) & 3.42(18)& 3.43(18)\\
\tableline
$\mu_{l,{\rm hel}}(N)$ [$\masyr$]  & -7.81(49) & -2.55(50) & -7.80(49)& -2.54(50)\\
$\mu_{l,{\rm hel}}(E)$ [$\masyr$]  & -4.43(48) & -4.67(49) & -4.44(48) & -4.65(49) \\
\tableline
$v_{l,{\rm pec}}(l)$ [$\kms$]  & -260(10)& -193(10) & -261(10)& -192(10)\\
$v_{l,{\rm pec}}(b)$ [$\kms$]  &    -3(9)&   54(10) &   -3(9) &   54(10) \\
\tableline
\end{tabular}
\end{table}

\begin{thebibliography}{}
\expandafter\ifx\csname natexlab\endcsname\relax\def\natexlab#1{#1}\fi
\providecommand{\url}[1]{\href{#1}{#1}}
\providecommand{\dodoi}[1]{doi:~\href{http://doi.org/#1}{\nolinkurl{#1}}}
\providecommand{\doeprint}[1]{\href{http://ascl.net/#1}{\nolinkurl{http://ascl.net/#1}}}
\providecommand{\doarXiv}[1]{\href{https://arxiv.org/abs/#1}{\nolinkurl{https://arxiv.org/abs/#1}}}

\bibitem[{{Abe} {et~al.}(2013){Abe}, {Airey}, {Barnard}, {Baudry}, {Botzler},
  {Douchin}, {Freeman}, {Larsen}, {Niemiec}, {Perrott}, {Philpott},
  {Rattenbury}, \& {Yock}}]{Abe.2013.A}
{Abe}, F., {Airey}, C., {Barnard}, E., {et~al.} 2013, \mnras, 431, 2975,
  \dodoi{10.1093/mnras/stt318}

\bibitem[{{Alard} \& {Lupton}(1998)}]{Alard.1998.A}
{Alard}, C., \& {Lupton}, R.~H. 1998, \apj, 503, 325, \dodoi{10.1086/305984}

\bibitem[{{Albrow} {et~al.}(2009){Albrow}, {Horne}, {Bramich}, {Fouqu{\'e}},
  {Miller}, {Beaulieu}, {Coutures}, {Menzies}, {Williams}, {Batista},
  {Bennett}, {Brillant}, {Cassan}, {Dieters}, {Prester}, {Donatowicz},
  {Greenhill}, {Kains}, {Kane}, {Kubas}, {Marquette}, {Pollard}, {Sahu},
  {Tsapras}, {Wambsganss}, \& {Zub}}]{Albrow.2009.A}
{Albrow}, M.~D., {Horne}, K., {Bramich}, D.~M., {et~al.} 2009, \mnras, 397,
  2099, \dodoi{10.1111/j.1365-2966.2009.15098.x}

\bibitem[{{Alcock} {et~al.}(2001){Alcock}, {Allsman}, {Alves}, {Axelrod},
  {Becker}, {Bennett}, {Cook}, {Drake}, {Freeman}, {Geha}, {Griest}, {Keller},
  {Lehner}, {Marshall}, {Minniti}, {Nelson}, {Peterson}, {Popowski}, {Pratt},
  {Quinn}, {Stubbs}, {Sutherland}, {Tomaney}, {Vandehei}, \&
  {Welch}}]{Alcock.2001.A}
{Alcock}, C., {Allsman}, R.~A., {Alves}, D.~R., {et~al.} 2001, \nat, 414, 617,
  \dodoi{10.1038/414617a}

\bibitem[{{An} {et~al.}(2002){An}, {Albrow}, {Beaulieu}, {Caldwell}, {DePoy},
  {Dominik}, {Gaudi}, {Gould}, {Greenhill}, {Hill}, {Kane}, {Martin},
  {Menzies}, {Pogge}, {Pollard}, {Sackett}, {Sahu}, {Vermaak}, {Watson}, \&
  {Williams}}]{An.2002.A}
{An}, J.~H., {Albrow}, M.~D., {Beaulieu}, J.-P., {et~al.} 2002, \apj, 572, 521,
  \dodoi{10.1086/340191}

\bibitem[{{Armstrong} \& {Bekki}(2018)}]{Armstrong.2018.A}
{Armstrong}, B., \& {Bekki}, K. 2018, \mnras, 480, L141,
  \dodoi{10.1093/mnrasl/sly143}

\bibitem[{{Batista} {et~al.}(2015){Batista}, {Beaulieu}, {Bennett}, {Gould},
  {Marquette}, {Fukui}, \& {Bhattacharya}}]{Batista.2015.A}
{Batista}, V., {Beaulieu}, J.-P., {Bennett}, D.~P., {et~al.} 2015, \apj, 808,
  170, \dodoi{10.1088/0004-637X/808/2/170}

\bibitem[{{Bennett} {et~al.}(2015){Bennett}, {Bhattacharya}, {Anderson},
  {Bond}, {Anderson}, {Barry}, {Batista}, {Beaulieu}, {DePoy}, {Dong}, {Gaudi},
  {Gilbert}, {Gould}, {Pfeifle}, {Pogge}, {Suzuki}, {Terry}, \&
  {Udalski}}]{Bennett.2015.A}
{Bennett}, D.~P., {Bhattacharya}, A., {Anderson}, J., {et~al.} 2015, \apj, 808,
  169, \dodoi{10.1088/0004-637X/808/2/169}

\bibitem[{{Bensby} {et~al.}(2013){Bensby}, {Yee}, {Feltzing}, {Johnson},
  {Gould}, {Cohen}, {Asplund}, {Mel{\'e}ndez}, {Lucatello}, {Han}, {Thompson},
  {Gal-Yam}, {Udalski}, {Bennett}, {Bond}, {Kohei}, {Sumi}, {Suzuki}, {Suzuki},
  {Takino}, {Tristram}, {Yamai}, \& {Yonehara}}]{Bensby.2013.A}
{Bensby}, T., {Yee}, J.~C., {Feltzing}, S., {et~al.} 2013, \aap, 549, A147,
  \dodoi{10.1051/0004-6361/201220678}

\bibitem[{{Bessell} \& {Brett}(1988)}]{Bessell.1988.A}
{Bessell}, M.~S., \& {Brett}, J.~M. 1988, \pasp, 100, 1134,
  \dodoi{10.1086/132281}

\bibitem[{{Bramich} {et~al.}(2008){Bramich}, {Vidrih}, {Wyrzykowski}, {Munn},
  {Lin}, {Evans}, {Smith}, {Belokurov}, {Gilmore}, {Zucker}, {Hewett},
  {Watkins}, {Faria}, {Fellhauer}, {Miknaitis}, {Bizyaev}, {Ivezi{\'c}},
  {Schneider}, {Snedden}, {Malanushenko}, {Malanushenko}, \&
  {Pan}}]{Bramich.2008.A}
{Bramich}, D.~M., {Vidrih}, S., {Wyrzykowski}, L., {et~al.} 2008, \mnras, 386,
  887, \dodoi{10.1111/j.1365-2966.2008.13053.x}

\bibitem[{{Calchi Novati} {et~al.}(2015{\natexlab{a}}){Calchi Novati}, {Gould},
  {Yee}, {Beichman}, {Bryden}, {Carey}, {Fausnaugh}, {Gaudi}, {Henderson},
  {Pogge}, {Shvartzvald}, {Wibking}, {Zhu}, {Spitzer Team}, {Udalski},
  {Poleski}, {Pawlak}, {Szyma{\'n}ski}, {Skowron}, {Mr{\'o}z}, {Koz{\l}owski},
  {Wyrzykowski}, {Pietrukowicz}, {Pietrzy{\'n}ski}, {Soszy{\'n}ski}, {Ulaczyk},
  \& {OGLE Group}}]{Calchi.2015.B}
{Calchi Novati}, S., {Gould}, A., {Yee}, J.~C., {et~al.} 2015{\natexlab{a}},
  \apj, 814, 92, \dodoi{10.1088/0004-637X/814/2/92}

\bibitem[{{Calchi Novati} {et~al.}(2015{\natexlab{b}}){Calchi Novati}, {Gould},
  {Udalski}, {Menzies}, {Bond}, {Shvartzvald}, {Street}, {Hundertmark},
  {Beichman}, {Yee}, {Carey}, {Poleski}, {Skowron}, {Koz{\l}owski}, {Mr{\'o}z},
  {Pietrukowicz}, {Pietrzy{\'n}ski}, {Szyma{\'n}ski}, {Soszy{\'n}ski},
  {Ulaczyk}, {Wyrzykowski}, {OGLE Collaboration}, {Albrow}, {Beaulieu},
  {Caldwell}, {Cassan}, {Coutures}, {Danielski}, {Dominis Prester},
  {Donatowicz}, {Lon{\v c}ari{\'c}}, {McDougall}, {Morales}, {Ranc}, {Zhu},
  {PLANET Collaboration}, {Abe}, {Barry}, {Bennett}, {Bhattacharya},
  {Fukunaga}, {Inayama}, {Koshimoto}, {Namba}, {Sumi}, {Suzuki}, {Tristram},
  {Wakiyama}, {Yonehara}, {The MOA Collaboration}, {Maoz}, {Kaspi},
  {Friedmann}, {Wise Group}, {Bachelet}, {Figuera Jaimes}, {Bramich},
  {Tsapras}, {Horne}, {Snodgrass}, {Wambsganss}, {Steele}, {Kains}, {RoboNet
  Collaboration}, {Bozza}, {Dominik}, {J{\o}rgensen}, {Alsubai}, {Ciceri},
  {D'Ago}, {Haugb{\o}lle}, {Hessman}, {Hinse}, {Juncher}, {Korhonen},
  {Mancini}, {Popovas}, {Rabus}, {Rahvar}, {Scarpetta}, {Schmidt}, {Skottfelt},
  {Southworth}, {Starkey}, {Surdej}, {Wertz}, {Zarucki}, {MiNDSTEp Consortium},
  {Gaudi}, {Pogge}, {DePoy}, \& {{$\mu$}FUN Collaboration}}]{Calchi.2015.A}
{Calchi Novati}, S., {Gould}, A., {Udalski}, A., {et~al.} 2015{\natexlab{b}},
  \apj, 804, 20, \dodoi{10.1088/0004-637X/804/1/20}

\bibitem[{{Camarillo} {et~al.}(2018){Camarillo}, {Dredger}, \&
  {Ratra}}]{Camarillo.2018.arXiv.A}
{Camarillo}, T., {Dredger}, P., \& {Ratra}, B. 2018, ArXiv e-prints.
\newblock \doarXiv{1805.01917}

\bibitem[{{Chabrier}(2003)}]{Chabrier.2003.A}
{Chabrier}, G. 2003, \pasp, 115, 763, \dodoi{10.1086/376392}

\bibitem[{{Chung} {et~al.}(2017){Chung}, {Zhu}, {Udalski}, {Lee}, {Ryu},
  {Jung}, {Shin}, {Yee}, {Hwang}, {Gould}, {and}, {Albrow}, {Cha}, {Han},
  {Kim}, {Kim}, {Kim}, {Kim}, {Lee}, {Park}, {Pogge}, {KMTNet Collaboration},
  {Poleski}, {Mr{\'o}z}, {Pietrukowicz}, {Skowron}, {Szyma{\'n}ski},
  {Soszy{\'n}ski}, {Koz{\l}owski}, {Ulaczyk}, {Pawlak}, {OGLE Collaboration},
  {Beichman}, {Bryden}, {Calchi Novati}, {Carey}, {Fausnaugh}, {Gaudi},
  {Henderson}, {Shvartzvald}, {Wibking}, \& {The Spitzer Team}}]{Chung.2017.A}
{Chung}, S.-J., {Zhu}, W., {Udalski}, A., {et~al.} 2017, \apj, 838, 154,
  \dodoi{10.3847/1538-4357/aa67fa}

\bibitem[{{Claret} \& {Bloemen}(2011)}]{Claret.2011.A}
{Claret}, A., \& {Bloemen}, S. 2011, \aap, 529, A75,
  \dodoi{10.1051/0004-6361/201116451}

\bibitem[{{Corsini} {et~al.}(2003){Corsini}, {Pizzella}, {Coccato}, \&
  {Bertola}}]{Corsini.2003.A}
{Corsini}, E.~M., {Pizzella}, A., {Coccato}, L., \& {Bertola}, F. 2003, \aap,
  408, 873, \dodoi{10.1051/0004-6361:20030951}

\bibitem[{{Cushing} {et~al.}(2009){Cushing}, {Looper}, {Burgasser},
  {Kirkpatrick}, {Faherty}, {Cruz}, {Sweet}, \& {Sanderson}}]{Cushing.2009.A}
{Cushing}, M.~C., {Looper}, D., {Burgasser}, A.~J., {et~al.} 2009, \apj, 696,
  986, \dodoi{10.1088/0004-637X/696/1/986}

\bibitem[{{Evans} {et~al.}(2016){Evans}, {Southworth}, {Maxted}, {Skottfelt},
  {Hundertmark}, {J{\o}rgensen}, {Dominik}, {Alsubai}, {Andersen}, {Bozza},
  {Bramich}, {Burgdorf}, {Ciceri}, {D'Ago}, {Figuera Jaimes}, {Gu},
  {Haugb{\o}lle}, {Hinse}, {Juncher}, {Kains}, {Kerins}, {Korhonen},
  {Kuffmeier}, {Mancini}, {Peixinho}, {Popovas}, {Rabus}, {Rahvar}, {Schmidt},
  {Snodgrass}, {Starkey}, {Surdej}, {Tronsgaard}, {von Essen}, {Wang}, \&
  {Wertz}}]{Evans.2016.A}
{Evans}, D.~F., {Southworth}, J., {Maxted}, P.~F.~L., {et~al.} 2016, \aap, 589,
  A58, \dodoi{10.1051/0004-6361/201527970}

\bibitem[{{Faherty} {et~al.}(2009){Faherty}, {Burgasser}, {Cruz}, {Shara},
  {Walter}, \& {Gelino}}]{Faherty.2009.A}
{Faherty}, J.~K., {Burgasser}, A.~J., {Cruz}, K.~L., {et~al.} 2009, \aj, 137,
  1, \dodoi{10.1088/0004-6256/137/1/1}

\bibitem[{{Gaia Collaboration} {et~al.}(2018){Gaia Collaboration}, {Brown},
  {Vallenari}, {Prusti}, {de Bruijne}, {Babusiaux}, \&
  {Bailer-Jones}}]{Gaia.2018arXiv.A}
{Gaia Collaboration}, {Brown}, A.~G.~A., {Vallenari}, A., {et~al.} 2018, ArXiv
  e-prints.
\newblock \doarXiv{1804.09365}

\bibitem[{{Gould}(1994{\natexlab{a}})}]{Gould.1994.B}
{Gould}, A. 1994{\natexlab{a}}, \apjl, 421, L71, \dodoi{10.1086/187190}

\bibitem[{{Gould}(1994{\natexlab{b}})}]{Gould.1994.A}
---. 1994{\natexlab{b}}, \apjl, 421, L75, \dodoi{10.1086/187191}

\bibitem[{{Gould}(2016)}]{Gould.2016.A}
---. 2016, Journal of Korean Astronomical Society, 49, 123,
  \dodoi{10.5303/JKAS.2016.49.4.123}

\bibitem[{{Gould} {et~al.}(2013){Gould}, {Carey}, \& {Yee}}]{Gould.2013.prop.A}
{Gould}, A., {Carey}, S., \& {Yee}, J. 2013, {Spitzer Microlens Planets and
  Parallaxes}, Spitzer Proposal

\bibitem[{{Gould} {et~al.}(2014){Gould}, {Carey}, \& {Yee}}]{Gould.2014.prop.A}
---. 2014, {Galactic Distribution of Planets from Spitzer Microlens
  Parallaxes}, Spitzer Proposal

\bibitem[{{Gould} {et~al.}(2016){Gould}, {Carey}, \& {Yee}}]{Gould.2016.prop.A}
---. 2016, {Galactic Distribution of Planets Spitzer Microlens Parallaxes},
  Spitzer Proposal

\bibitem[{{Gould} {et~al.}(2010){Gould}, {Dong}, {Bennett}, {Bond}, {Udalski},
  \& {Kozlowski}}]{Gould.2010.A}
{Gould}, A., {Dong}, S., {Bennett}, D.~P., {et~al.} 2010, \apj, 710, 1800,
  \dodoi{10.1088/0004-637X/710/2/1800}

\bibitem[{{Gould} \& {Loeb}(1992)}]{Gould.1992.A}
{Gould}, A., \& {Loeb}, A. 1992, \apj, 396, 104, \dodoi{10.1086/171700}

\bibitem[{{Gould} {et~al.}(2015{\natexlab{a}}){Gould}, {Yee}, \&
  {Carey}}]{Gould.2015.prop.A}
{Gould}, A., {Yee}, J., \& {Carey}, S. 2015{\natexlab{a}}, {Galactic
  Distribution of Planets From High-Magnification Microlensing Events}, Spitzer
  Proposal

\bibitem[{{Gould} {et~al.}(2015{\natexlab{b}}){Gould}, {Yee}, \&
  {Carey}}]{Gould.2015.prop.B}
---. 2015{\natexlab{b}}, {Degeneracy Breaking for K2 Microlens Parallaxes},
  Spitzer Proposal

\bibitem[{{Gould} \& {Yee}(2012)}]{Gould.2012.A}
{Gould}, A., \& {Yee}, J.~C. 2012, \apjl, 755, L17,
  \dodoi{10.1088/2041-8205/755/1/L17}

\bibitem[{{Gould} \& {Yee}(2014)}]{Gould.2014.B}
---. 2014, \apj, 784, 64, \dodoi{10.1088/0004-637X/784/1/64}

\bibitem[{{Griest} \& {Safizadeh}(1998)}]{Griest.1998.A}
{Griest}, K., \& {Safizadeh}, N. 1998, \apj, 500, 37, \dodoi{10.1086/305729}

\bibitem[{{Han} {et~al.}(2005){Han}, {Gaudi}, {An}, \& {Gould}}]{Han.2005.A}
{Han}, C., {Gaudi}, B.~S., {An}, J.~H., \& {Gould}, A. 2005, \apj, 618, 962,
  \dodoi{10.1086/426115}

\bibitem[{{Henderson} \& {Shvartzvald}(2016)}]{Henderson.2016.A}
{Henderson}, C.~B., \& {Shvartzvald}, Y. 2016, \aj, 152, 96,
  \dodoi{10.3847/0004-6256/152/4/96}

\bibitem[{{Hog} {et~al.}(1995){Hog}, {Novikov}, \& {Polnarev}}]{Hog.1995.A}
{Hog}, E., {Novikov}, I.~D., \& {Polnarev}, A.~G. 1995, \aap, 294, 287

\bibitem[{{Jiang} {et~al.}(2004){Jiang}, {DePoy}, {Gal-Yam}, {Gaudi}, {Gould},
  {Han}, {Lipkin}, {Maoz}, {Ofek}, {Park}, {Pogge}, {MuFun Collaboration},
  {Udalski}, {Kubiak}, {Szyma{\'n}ski}, {Szewczyk}, {{\.Z}ebru{\'n}},
  {Wyrzykowski}, {Soszy{\'n}ski}, {Pietrzy{\'n}ski}, {OGLE Collaboration},
  {Albrow}, {Beaulieu}, {Caldwell}, {Cassan}, {Coutures}, {Dominik},
  {Donatowicz}, {Fouqu{\'e}}, {Greenhill}, {Hill}, {Horne}, {J{\o}rgensen},
  {J{\o}rgensen}, {Kane}, {Kubas}, {Martin}, {Menzies}, {Pollard}, {Sahu},
  {Wambsganss}, {Watson}, {Williams}, \& {PLANET Collaboration}}]{Jiang.2004.A}
{Jiang}, G., {DePoy}, D.~L., {Gal-Yam}, A., {et~al.} 2004, \apj, 617, 1307,
  \dodoi{10.1086/425678}

\bibitem[{{Kervella} {et~al.}(2004){Kervella}, {Th{\'e}venin}, {Di Folco}, \&
  {S{\'e}gransan}}]{Kervella.2004.A}
{Kervella}, P., {Th{\'e}venin}, F., {Di Folco}, E., \& {S{\'e}gransan}, D.
  2004, \aap, 426, 297, \dodoi{10.1051/0004-6361:20035930}

\bibitem[{{Kim} {et~al.}(2016){Kim}, {Lee}, {Park}, {Kim}, {Cha}, {Lee}, {Han},
  {Chun}, \& {Yuk}}]{Kim.2016.A}
{Kim}, S.-L., {Lee}, C.-U., {Park}, B.-G., {et~al.} 2016, Journal of Korean
  Astronomical Society, 49, 37, \dodoi{10.5303/JKAS.2016.49.1.037}

\bibitem[{{Kroupa}(2001)}]{Kroupa.2001.A}
{Kroupa}, P. 2001, \mnras, 322, 231, \dodoi{10.1046/j.1365-8711.2001.04022.x}

\bibitem[{{Marigo} {et~al.}(2017){Marigo}, {Girardi}, {Bressan}, {Rosenfield},
  {Aringer}, {Chen}, {Dussin}, {Nanni}, {Pastorelli}, {Rodrigues}, {Trabucchi},
  {Bladh}, {Dalcanton}, {Groenewegen}, {Montalb{\'a}n}, \&
  {Wood}}]{Marigo.2017.A}
{Marigo}, P., {Girardi}, L., {Bressan}, A., {et~al.} 2017, \apj, 835, 77,
  \dodoi{10.3847/1538-4357/835/1/77}

\bibitem[{{Miyamoto} \& {Yoshii}(1995)}]{Miyamoto.1995.A}
{Miyamoto}, M., \& {Yoshii}, Y. 1995, \aj, 110, 1427, \dodoi{10.1086/117616}

\bibitem[{{Mr{\'o}z} {et~al.}(2018){Mr{\'o}z}, {Ryu}, {Skowron}, {Udalski},
  {Gould}, {Szyma{\'n}ski}, {Soszy{\'n}ski}, {Poleski}, {Pietrukowicz},
  {Koz{\l}owski}, {Pawlak}, {Ulaczyk}, {OGLE Collaboration}, {Albrow}, {Chung},
  {Jung}, {Han}, {Hwang}, {Shin}, {Yee}, {Zhu}, {Cha}, {Kim}, {Kim}, {Kim},
  {Lee}, {Lee}, {Lee}, {Park}, {Pogge}, \& {KMTNet
  Collaboration}}]{Mroz.2018.A}
{Mr{\'o}z}, P., {Ryu}, Y.-H., {Skowron}, J., {et~al.} 2018, \aj, 155, 121,
  \dodoi{10.3847/1538-3881/aaaae9}

\bibitem[{{Nataf} {et~al.}(2013){Nataf}, {Gould}, {Fouqu{\'e}}, {Gonzalez},
  {Johnson}, {Skowron}, {Udalski}, {Szyma{\'n}ski}, {Kubiak},
  {Pietrzy{\'n}ski}, {Soszy{\'n}ski}, {Ulaczyk}, {Wyrzykowski}, \&
  {Poleski}}]{Nataf.2013.A}
{Nataf}, D.~M., {Gould}, A., {Fouqu{\'e}}, P., {et~al.} 2013, \apj, 769, 88,
  \dodoi{10.1088/0004-637X/769/2/88}

\bibitem[{{Paczy{\'n}ski}(1986)}]{Paczynski.1986.A}
{Paczy{\'n}ski}, B. 1986, \apj, 304, 1, \dodoi{10.1086/164140}

\bibitem[{{Pizzella} {et~al.}(2004){Pizzella}, {Corsini}, {Vega Beltr{\'a}n},
  \& {Bertola}}]{Pizzella.2004.A}
{Pizzella}, A., {Corsini}, E.~M., {Vega Beltr{\'a}n}, J.~C., \& {Bertola}, F.
  2004, \aap, 424, 447, \dodoi{10.1051/0004-6361:20047183}

\bibitem[{{Pizzella} {et~al.}(2014){Pizzella}, {Morelli}, {Corsini}, {Dalla
  Bont{\`a}}, {Coccato}, \& {Sanjana}}]{Pizzella.2014.A}
{Pizzella}, A., {Morelli}, L., {Corsini}, E.~M., {et~al.} 2014, \aap, 570, A79,
  \dodoi{10.1051/0004-6361/201424746}

\bibitem[{{Refsdal}(1966)}]{Refsdal.1966.A}
{Refsdal}, S. 1966, \mnras, 134, 315, \dodoi{10.1093/mnras/134.3.315}

\bibitem[{{Rubin}(1994)}]{Rubin.1994.A}
{Rubin}, V.~C. 1994, \aj, 108, 456, \dodoi{10.1086/117083}

\bibitem[{{Schechter} {et~al.}(1993){Schechter}, {Mateo}, \&
  {Saha}}]{Schechter.1993.A}
{Schechter}, P.~L., {Mateo}, M., \& {Saha}, A. 1993, \pasp, 105, 1342,
  \dodoi{10.1086/133316}

\bibitem[{{Sch{\"o}nrich} {et~al.}(2010){Sch{\"o}nrich}, {Binney}, \&
  {Dehnen}}]{Schonrich.2010.A}
{Sch{\"o}nrich}, R., {Binney}, J., \& {Dehnen}, W. 2010, \mnras, 403, 1829,
  \dodoi{10.1111/j.1365-2966.2010.16253.x}

\bibitem[{{Shin} {et~al.}(2018){Shin}, {Udalski}, {Yee}, {Calchi Novati},
  {Christie}, {Poleski}, {Mr{\'o}z}, {Skowron}, {Szyma{\'n}ski},
  {Soszy{\'n}ski}, {Pietrukowicz}, {Koz{\l}owski}, {Ulaczyk}, {Pawlak},
  {Natusch}, {Pogge}, {Gould}, {Han}, {Albrow}, {Chung}, {Hwang}, {Ryu},
  {Jung}, {Zhu}, {Lee}, {Cha}, {Kim}, {Kim}, {Kim}, {Lee}, {Lee}, {Park},
  {Beichman}, {Bryden}, {Carey}, {Gaudi}, {Henderson}, \&
  {Shvartzvald}}]{Shin.2018.arXiv.A}
{Shin}, I.-G., {Udalski}, A., {Yee}, J.~C., {et~al.} 2018, AJ submitted,
  arXiv:1801.00169.
\newblock \doarXiv{1801.00169}

\bibitem[{{Shvartzvald} {et~al.}(2017){Shvartzvald}, {Yee}, {Calchi Novati},
  {Gould}, {Lee}, {Beichman}, {Bryden}, {Carey}, {Gaudi}, {Henderson}, {Zhu},
  {Spitzer Team}, {Albrow}, {Cha}, {Chung}, {Han}, {Hwang}, {Jung}, {Kim},
  {Kim}, {Kim}, {Lee}, {Park}, {Pogge}, {Ryu}, {Shin}, \& {KMTNet
  Group}}]{Shvartzvald.2017.B}
{Shvartzvald}, Y., {Yee}, J.~C., {Calchi Novati}, S., {et~al.} 2017, \apjl,
  840, L3, \dodoi{10.3847/2041-8213/aa6d09}

\bibitem[{{Simpson}(1740)}]{Simpson.1740.A}
{Simpson}, T. 1740, printed by H. Woodfall, jun. for J. Nourse, Section 6, pp.
  81-86

\bibitem[{{Skottfelt} {et~al.}(2015){Skottfelt}, {Bramich}, {Hundertmark},
  {J{\o}rgensen}, {Michaelsen}, {Kj{\ae}rgaard}, {Southworth}, {S{\o}rensen},
  {Andersen}, {Andersen}, {Christensen-Dalsgaard}, {Frandsen}, {Grundahl},
  {Harps{\o}e}, {Kjeldsen}, \& {Pall{\'e}}}]{Skottfelt.2015.A}
{Skottfelt}, J., {Bramich}, D.~M., {Hundertmark}, M., {et~al.} 2015, \aap, 574,
  A54, \dodoi{10.1051/0004-6361/201425260}

\bibitem[{{Skowron} \& {Gould}(2012)}]{Skowron.2012.arXiv.A}
{Skowron}, J., \& {Gould}, A. 2012, ArXiv e-prints.
\newblock \doarXiv{1203.1034}

\bibitem[{{Udalski}(2003)}]{Udalski.2003.A}
{Udalski}, A. 2003, \actaa, 53, 291

\bibitem[{{Udalski} {et~al.}(1994){Udalski}, {Szymanski}, {Kaluzny}, {Kubiak},
  {Mateo}, {Krzeminski}, \& {Paczynski}}]{Udalski.1994.A}
{Udalski}, A., {Szymanski}, M., {Kaluzny}, J., {et~al.} 1994, \actaa, 44, 227

\bibitem[{{Udalski} {et~al.}(2015){Udalski}, {Szyma{\'n}ski}, \&
  {Szyma{\'n}ski}}]{Udalski.2015.B}
{Udalski}, A., {Szyma{\'n}ski}, M.~K., \& {Szyma{\'n}ski}, G. 2015, \actaa, 65,
  1.
\newblock \doarXiv{1504.05966}

\bibitem[{{Walker}(1995)}]{Walker.1995.A}
{Walker}, M.~A. 1995, \apj, 453, 37, \dodoi{10.1086/176367}

\bibitem[{{Wozniak}(2000)}]{Wozniak.2000.A}
{Wozniak}, P.~R. 2000, \actaa, 50, 421

\bibitem[{{Yee} {et~al.}(2015){Yee}, {Gould}, {Beichman}, {Calchi Novati},
  {Carey}, {Gaudi}, {Henderson}, {Nataf}, {Penny}, {Shvartzvald}, \&
  {Zhu}}]{Yee.2015.A}
{Yee}, J.~C., {Gould}, A., {Beichman}, C., {et~al.} 2015, \apj, 810, 155,
  \dodoi{10.1088/0004-637X/810/2/155}

\bibitem[{{Zhang} {et~al.}(2017){Zhang}, {Pinfield}, {G{\'a}lvez-Ortiz},
  {Burningham}, {Lodieu}, {Marocco}, {Burgasser}, {Day-Jones}, {Allard},
  {Jones}, {Homeier}, {Gomes}, \& {Smart}}]{Zhang.2017.A}
{Zhang}, Z.~H., {Pinfield}, D.~J., {G{\'a}lvez-Ortiz}, M.~C., {et~al.} 2017,
  \mnras, 464, 3040, \dodoi{10.1093/mnras/stw2438}

\bibitem[{{Zhu} {et~al.}(2016){Zhu}, {Calchi Novati}, {Gould}, {Udalski},
  {Han}, {Shvartzvald}, {Ranc}, {J{\o}rgensen}, {Poleski}, {Bozza}, {Beichman},
  {Bryden}, {Carey}, {Gaudi}, {Henderson}, {Pogge}, {Porritt}, {Wibking},
  {Yee}, {SPITZER Team}, {Pawlak}, {Szyma{\'n}ski}, {Skowron}, {Mr{\'o}z},
  {Koz{\l}owski}, {Wyrzykowski}, {Pietrukowicz}, {Pietrzy{\'n}ski},
  {Soszy{\'n}ski}, {Ulaczyk}, {OGLE Group}, {Choi}, {Park}, {Jung}, {Shin},
  {Albrow}, {Park}, {Kim}, {Lee}, {Cha}, {Kim}, {Lee}, {KMTNET Group},
  {Friedmann}, {Kaspi}, {Maoz}, {WISE Group}, {Hundertmark}, {Street},
  {Tsapras}, {Bramich}, {Cassan}, {Dominik}, {Bachelet}, {Dong}, {Figuera
  Jaimes}, {Horne}, {Mao}, {Menzies}, {Schmidt}, {Snodgrass}, {Steele},
  {Wambsganss}, {RoboNeT Team}, {Skottfelt}, {Andersen}, {Burgdorf}, {Ciceri},
  {D'Ago}, {Evans}, {Gu}, {Hinse}, {Kerins}, {Korhonen}, {Kuffmeier},
  {Mancini}, {Peixinho}, {Popovas}, {Rabus}, {Rahvar}, {Tronsgaard},
  {Scarpetta}, {Southworth}, {Surdej}, {von Essen}, {Wang}, {Wertz}, \&
  {MiNDSTEP Group}}]{Zhu.2016.A}
{Zhu}, W., {Calchi Novati}, S., {Gould}, A., {et~al.} 2016, \apj, 825, 60,
  \dodoi{10.3847/0004-637X/825/1/60}

\end{thebibliography}
\end{document}